\documentclass[aps,prb,amsmath,amssymb,reprint,superscriptaddress,preprintnumbers,showpacs,intlimits,longbibliography]{revtex4-1}
\usepackage{bm,latexsym,mathrsfs,enumerate,color}
\usepackage[mathcal]{euscript}
\usepackage[breaklinks=true,unicode=true,urlcolor = blue,colorlinks = true,citecolor = blue,linkcolor = blue]{hyperref}
\usepackage{graphicx}
\usepackage{adjustbox}
\usepackage{array,ragged2e}
\usepackage[cal=boondoxo]{mathalfa}
\usepackage{lastpage}
\DeclareGraphicsExtensions{.pdf,.svg,.eps,.ps,.png,.jpg,.jpeg}

\graphicspath{{../figs/}}
\renewcommand{\vec}[1]{\bm{#1}}
\DeclareMathOperator{\sech}{sech}

\bibpunct{[}{]}{,}{n}{}{}

\begin{document}
%
%
\title{Spin eigen-excitations of an antiferromagnetic skyrmion}

\author{Volodymyr P. Kravchuk}
\email[Corresponding author: ]{v.kravchuk@ifw-dresden.de}
\affiliation{Leibniz-Institut f{\"u}r Festk{\"o}rper- und Werkstoffforschung, IFW Dresden, D-01171 Dresden, Germany}
\affiliation{Bogolyubov Institute for Theoretical Physics of National Academy of Sciences of Ukraine, 03680 Kyiv, Ukraine}

\author{Olena Gomonay}
\email{ogomonay@uni-mainz.de}
\affiliation{Institut f{\"u}r Physik, Johannes Gutenberg-Universit{\"a}t Mainz, D-55128 Mainz, Germany}

\author{Denis D. Sheka}
\email{sheka@knu.ua}
\affiliation{Taras Shevchenko National University of Kyiv, 01601 Kyiv, Ukraine}

\author{Davi R. Rodrigues}
\email{davrodri@uni-mainz.de}
\affiliation{Institut f{\"u}r Physik, Johannes Gutenberg-Universit{\"a}t Mainz, D-55128 Mainz, Germany}
 
\author{Karin~Everschor-Sitte}
\email{kaeversc@uni-mainz.de}
\affiliation{Institut f{\"u}r Physik, Johannes Gutenberg-Universit{\"a}t Mainz, D-55128 Mainz, Germany}

\author{Jairo Sinova}
\email{sinova@uni-mainz.de}
\affiliation{Institut f{\"u}r Physik, Johannes Gutenberg-Universit{\"a}t Mainz, D-55128 Mainz, Germany}

\author{Jeroen~van~den~Brink}
\email{j.van.den.brink@ifw-dresden.de}
\affiliation{Leibniz-Institut f{\"u}r Festk{\"o}rper- und Werkstoffforschung, IFW Dresden, D-01171 Dresden, Germany}
\affiliation{Institute for Theoretical Physics, TU Dresden, 01069 Dresden, Germany}
\affiliation{Department of Physics, Washington University, St. Louis, MO 63130, USA}

\author{Yuri Gaididei}
\email{ybg@bitp.kiev.ua}
\affiliation{Bogolyubov Institute for Theoretical Physics of National Academy of Sciences of Ukraine, 03680 Kyiv, Ukraine}


%
%
%
%
\begin{abstract}
We theoretically predict and classify the localized modes of a skyrmion in a collinear uniaxial antiferromagnet and discuss how they can be excited.  
As a central result we find two branches of skyrmion eigenmodes with distinct physical properties characterized by being low or high energy excitations.
The frequency dependence of the low-energy modes scales as $R_0^{-2}$ for skyrmions with large radius $R_0$. 
Furthermore, we predict localized high-energy eigenmodes, which have no direct ferromagnetic counterpart. 
Except for the breathing mode, we find that all localized antiferromagnet skyrmion modes, both in the low and high-energy branch, are doubly degenerated in the absence of a magnetic field and split otherwise. 
We explain our numerical results for the low-energy modes within a string model representing the skyrmion boundary.

\end{abstract}
%

\maketitle


\section{Introduction}
Antiferromagnets (AFMs) recently have become promising as active elements in spintronic devices as they have faster dynamics and are insensitive to magnetic fields, for reviews see Refs.~ \onlinecite{MacDonald2011, Gomonay14, Gomonay2017}.  New ways to manipulate AFM textures have been predicted and experimental methods are rapidly proceeding towards imaging and controlling them \cite{Wadley2016,Zelezny2018}.
For example, AFM domain walls have been observed and manipulated already \cite{Kim2017}. Still several challenges remain due to the difficulty to access directly the AFM order parameter.

Recently, AFM skyrmions have attracted a lot of attention due to their interesting properties such as the absence of a skyrmion Hall effect   \cite{Barker16,Zhang16a,Gobel2017} and the presence of the topological spin Hall effect \cite{Gobel2017}.
While these localized metastable objects have been predicted several years ago \cite{Bogdanov02a}, but so far mostly their static properties as well as their translational motion has been studied \cite{Barker16,Zhang16a,Velkov16,Gomonay2018,Smejkal2017b}.

In this work we analyze the excitation modes of AFM skyrmions and thereby predict novel approaches for their experimental observation.
We find two branches of the localized eigenmodes and analyze their structure as a function of the magnetic field, skyrmion radius, and strength of Dzyaloshinskii-Moria interaction (DMI). For the interpretation of the low-frequency dynamics of the skyrmion boundary we develop  a string-model of the AFM domain wall.

This work is structured as follows. After introducing the AFM model in Sec.~\ref{sec:Model} we review the static skyrmion solution in Sec.~\ref{sec:static} focusing on the large skyrmion limit. In Sec.~\ref{sec:magnons} we formulate the eigen-value problem for the AFM skyrmion and discuss its localized eigenmodes based on the result of numerical calculations. In Sec.~\ref{sec:circ-domain-wall} we introduce an effective model of the AFM domain wall string and apply it to interpret of the low-frequency branch of the spectrum in the large skyrmion limit. The main results of the paper are summarized in the last~Sec.~\ref{sec:concl}.

\section{Model of an antiferromagnet}
\label{sec:Model}

\begin{figure}
	\includegraphics[width=\columnwidth]{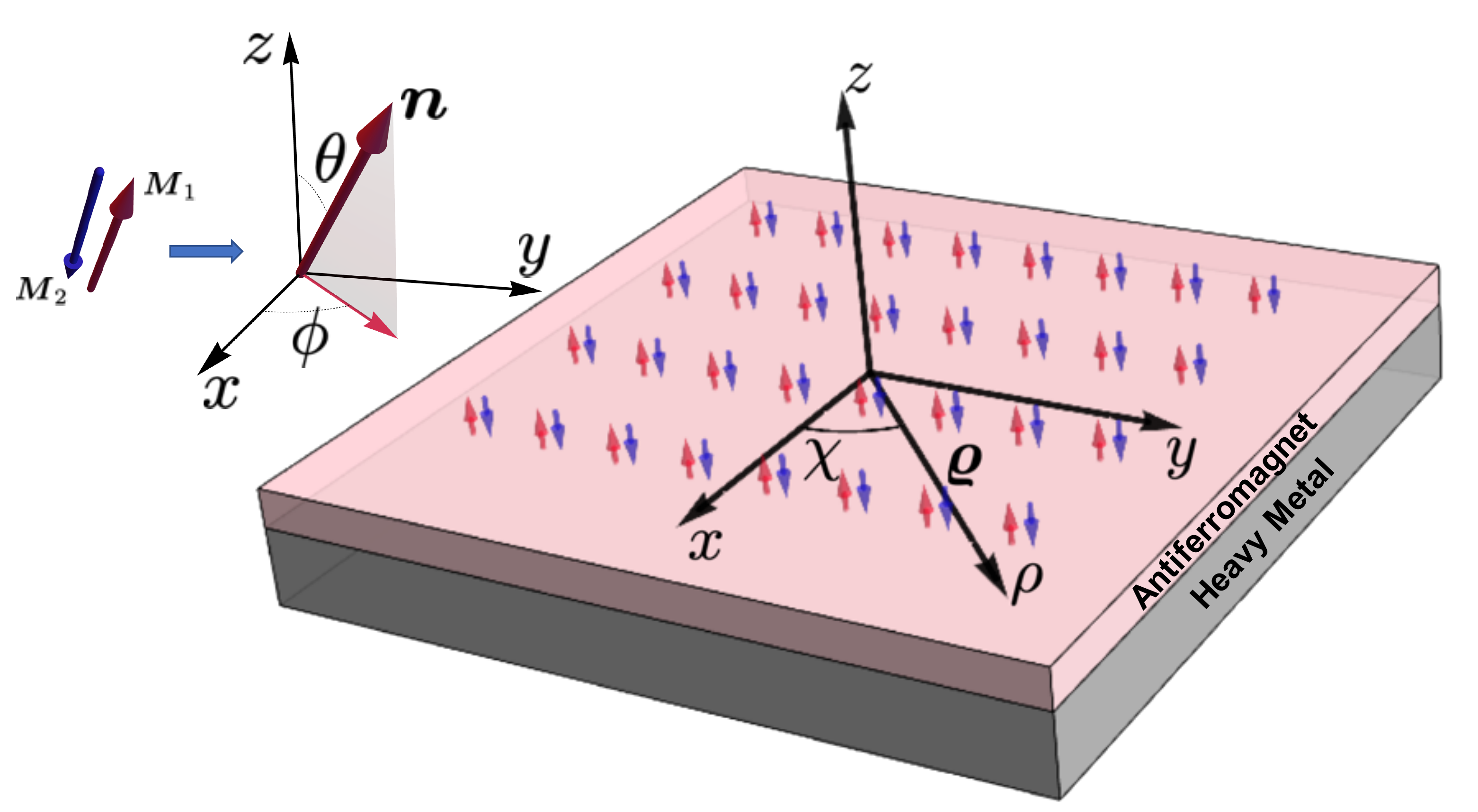}
	\caption{(Color online) \textbf{Sketch of setup} a thin film uniaxial collinear AFM with two magnetic sublattices  $\vec{M}_1$ and $\vec{M}_2$  on top of a heavy metal substrate. Inset: parametrization of the N\'eel vector in terms of spherical coordinates $\theta$ and $\phi$.
	}\label{fig:setup}
\end{figure}

We consider a thin film uniaxial collinear AFM with two magnetic sublattices  $\vec{M}_1$ and $\vec{M}_2$ (with $|\vec{M}_1|=|\vec{M}_2|=M_s$) which are antiparallel and fully compensate each other in the equilibrium state. 
We chose the normal $\hat {\vec{z}}$ of the film parallel to the magnetic easy axis, see Fig.~\ref{fig:setup}. The coupling of the magnetic film to a heavy metal substrate breakes the inversion symmetry and creates an interfacial DMI whose strength $D$ can be controlled by the AFM thickness.
The thickness of the AFM film is assumed to be small enough to exclude any inhomogeneities along the vertical direction.
The order parameters of this AFM are given by the N\'eel vector $\vec{n}=(\vec{M}_1-\vec{M}_2)/(2M_s)$ and the magnetization
$\vec{m}=(\vec{M}_1+\vec{M}_2)/(2M_s)$. 

In the following we will consider the case of a strong exchange field $H_\mathrm{ex}$ \footnote{The exchange interaction which keeps the magnetization vectors on the different sublattices antiparallel, is parametrized by the value of the $H_\mathrm{ex}=\mathfrak{A}/M_s$, where $\mathfrak{A}$ is the constant of the uniform exchange energy with density $\mathcal{E}_{\rm ex}^{\textsc u}=\mathfrak{A}\vec{m}^2$.} 
acting between the magnetic sublattices. It is large compared to the anisotropy field  $H_\mathrm{an}$, i.e. $H_\mathrm{an}/H_\mathrm{ex} \equiv \xi^2 \ll 1$. In this case the magnetization $\vec{m}$ is small (i.e.\ $|\vec{m}|\ll1$ and $|\vec{n}|\approx 1$), and the state of the AFM is determined solely by the spatial and time dependent N\'eel vector. 
The effective energy of such an AFM is given by 
\begin{equation}\label{eq:E-tot}
E= L\int \mathrm{d} ^2 x\left[A(\partial_i \vec n \cdot \partial_i \vec n)+M_sH_\mathrm{an}\left(1-n_z^2\right)+ \mathcal{E}_\textsc{dmi} \right],
\end{equation}
where $L$ is the thickness of the film, $A$ is the exchange stiffness, and $\mathcal{E}_\textsc{dmi} =D ( n_z\vec{\nabla}\cdot\vec{n}-\vec{n}\cdot\vec{\nabla}n_z )$ is the interfacial DMI, typical for AFMs of $C_{nv}$ symmetry classes \cite{Bogdanov02a}. An example of material that can be described within such a model is the recently synthesed compensated Heusler compound Mn$_2$Ru$_x$Ga \cite{Thiyagarajah2015}. Note that Einstein  summation rule is used in Eq.~\eqref{eq:E-tot}.

The dynamics of the N\'eel vector in the presence of an external magnetic field 
can be effectively described within the Lagrange formalism \cite{Baryakhtar1980, Kosevich1990, Gomonay10}. The Lagrangian in rescaled parameters (see Table \ref{tbl:units}) is given by
\begin{subequations} \label{eq:L}
\begin{align} \label{eq:L-1}
&\mathscr{L}=\left[\dot{\vec{n}}-(\vec{h}\times\vec{n})\right]^2-\mathcal{W},\\
\label{eq:L-2} 
&\mathcal{W} = \partial_i \vec n \cdot \partial_i \vec n + 1-n_z^2 +d ( n_z \partial_i n_i-n_i \partial_i n_z ) 
\end{align}
\end{subequations}
and is supplemented by the constraint $|\vec{n}|=1$. Here $d=D/\sqrt{AH_\mathrm{an}M_s}$  is the DMI strength and $\vec{h}=\vec{H}/H_\mathrm{sf}$  represents the rescaled magnetic field, where  $H_\mathrm{sf}=\sqrt{H_\mathrm{an}H_\mathrm{ex}}$ is the spin-flop field.
Note that a small magnetization arises either due to the presence of a magnetic field or is induced by the dynamics of the N\'eel vector
\begin{equation}\label{eq:m-dyn}
\vec{m}=\xi\left[\dot{\vec{n}}\times\vec{n}+\vec{n}\times\vec{h}\times\vec{n}\right].
\end{equation}
The above model, Eqs.~\eqref{eq:L},\eqref{eq:m-dyn}, is controlled by two dimensionless parameters, namely, the DMI constant $d$ and the magnetic field $\vec{h}$. The overdot indicates the derivative with respect to the dimensionless time $\tau=t\, \omega_\mathrm{AFMR}$, where $\omega_\mathrm{AFMR}=\gamma \sqrt{H_\mathrm{an}H_\mathrm{ex}}$ is the frequency of AFM resonance, $\gamma$ is gyromagnetic ratio. Note also that  the spatial  derivates are now measured with respect to the dimensionless length $r/\ell$
where $\ell$ is in units of the magnetic length $\ell=\sqrt{A/(H_\mathrm{an}M_s)}$ which describes the width of a plane domain wall, see Tab.~\ref{tbl:units}.
The small parameter $\xi$ can be associated with the value of the static magnetization, when the applied magnetic field is equal to the spin-flop field ($h=1$), or with the dynamical magnetization when the N\'eel vector lying within $xy$-plane precesses with the AFM resonance frequency, see Eq.~\eqref{eq:m-dyn}.
Despite the magnetization is small in antiferromagnetic systems, we show below that it can be used for resonant excitations of some eigenmodes by means of an ac magnetic field.
 
In the following we set the magnetic field parallel to the easy-axis, $\vec{h}=h\hat{\vec{z}}$.
As $|\vec{n}|=1$, we use a parametrization in spherical angular coordinates 
$\vec{n}=\sin\theta\, (\cos\phi \, \hat{\vec{x}}+\sin\phi\, \hat{\vec{y}})+ \cos\theta \, \hat{\vec{z}}$ leading to 
\begin{equation}\label{eq:L-angles}
\mathscr{L}=\dot{\theta}^2+(\dot{\phi}-h)^2 \sin^2\theta-\mathcal{W}[\theta, \phi].
\end{equation}
%

In the next sections, we will first review the static solution of an AFM skyrmion based on which we will then derive its excitations.

\begin{table*}
	\renewcommand{\arraystretch}{1.3}	
	\begin{tabular}{ll|l|l|l}
			\hline\hline
		Notation & Dimensionless quantity & Unit of \hfill \break measurement & Physical meaning&Typical value\\ \hline
		$\vec{h}=\vec{H}/H_\mathrm{sf}$  & Magnetic field & $H_\mathrm{sf}=\sqrt{H_\mathrm{an}H_\mathrm{ex}}$ & Spin-flop field &26 T\\ 
		$\tau=t \, \omega_\mathrm{AFMR}$ & Time  & $\omega_\mathrm{AFMR}=\gamma\sqrt{H_\mathrm{an}H_\mathrm{ex}}$ & Frequency of the uniform AFM resonance& $4.6\cdot 10^{12}$ sec$^{-1}$\\
		$\vec{\varrho}=\vec{r}/\ell$ & Length   & $\ell=\sqrt{A/(H_\mathrm{an}M_s)}$ & Width of a plane domain wall &6 nm\\
		$d=4 \delta/\pi = D/D_\mathrm{DW}$ & DMI strength  & $D_\mathrm{DW}=\sqrt{AH_\mathrm{an}M_s}$ & Domain wall energy density&8 mJ/m$^2$\\
		$\vec{m}=\vec{M}/M_0$ & Magnetization   & $M_0=2M_s$ & Saturation magnetization &1 MA/m
		\\  
		\hline
		$\xi=\sqrt{H_\mathrm{an}/H_\mathrm{ex}}$&Expansion parameter & &&  0.002\\
		\hline\hline
	\end{tabular}	
	\caption{Units of measurement used in this paper. Typical values  in the last column correspond to the Heusler compound Mn$_2$Ru$_x$Ga near its compensation point \cite{Siewierska2017, Fowley2018}. 
	}\label{tbl:units}
\end{table*}


\section{Static Skyrmion solution}
\label{sec:static}
An AFM skyrmion is a meta-stable state embedded in a homogeneous collinear AFM. In a previous work \cite{Bogdanov02a}, it has been shown that a homogenous collinear phase with the N\'eel vector oriented parallel to the easy axis ($\vec{n} || \hat{\vec{z}}$) of the AFM model in Eq.~\eqref{eq:L-2} requires 
\begin{equation}
\label{eq:exits}
h^2+\delta^2 <1 \quad \textrm{with} \quad \delta=\frac{\pi}{4} d.
\end{equation}
Here we have introduced the renormalized DMI constant $\delta$.
We denote the critical value where the AFM phase and the skyrmion solution become unstable as $h_c= \sqrt{1-\delta^2}$. As a result of instability, a modulated periodical structure (spin-flop phase) is developing if $\delta>0$ ($\delta=0$), see Ref.~\cite{Bogdanov02a} for details.

\begin{figure}
	\includegraphics[width=\columnwidth]{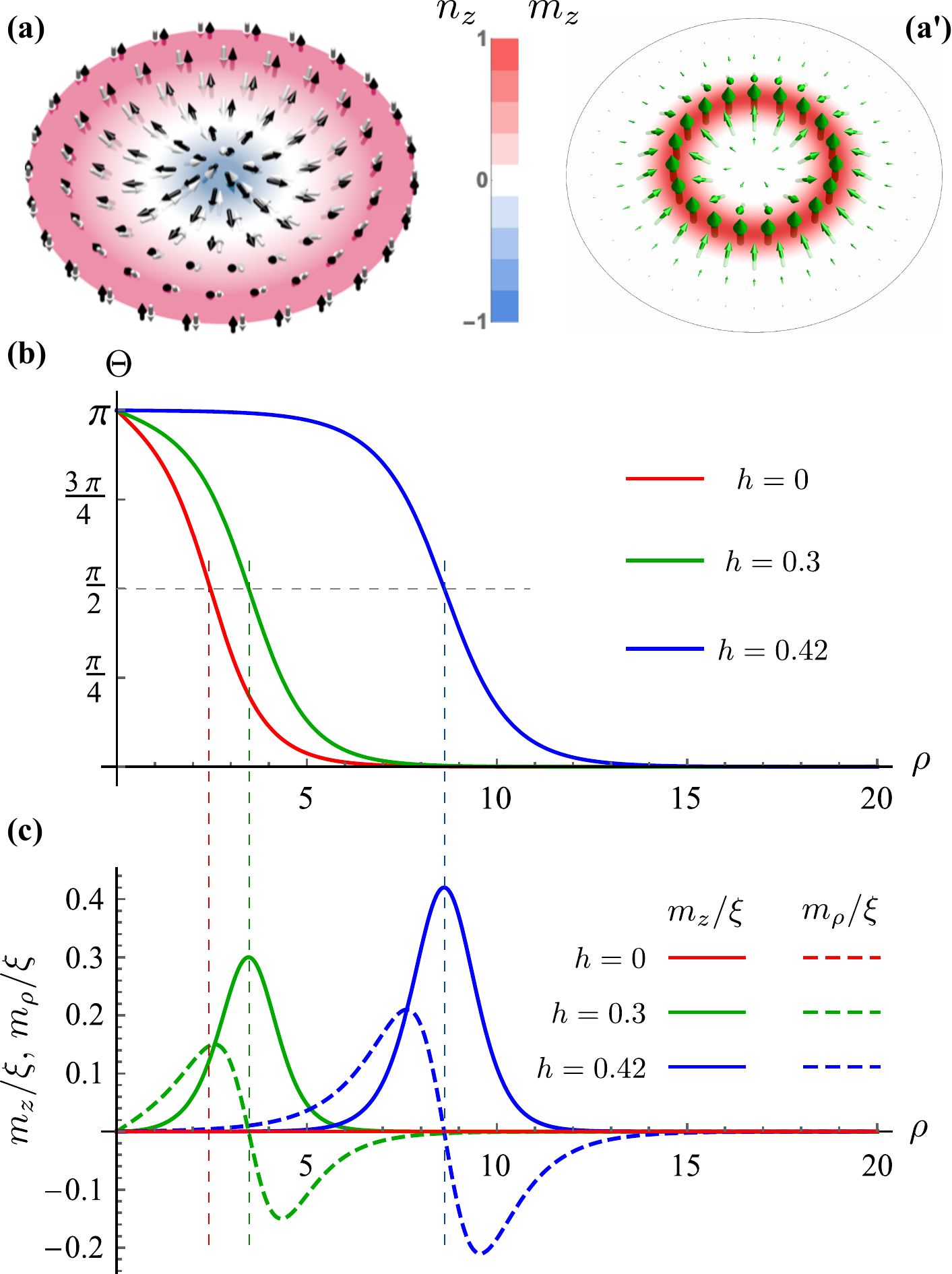}
	\caption{\textbf{(Color online) Structure of the AFM skyrmion.} (a) Distribution of magnetic moments, black and white arrows correspond to the different sublattices and the color scheme codes the perpendicular component $n_z$. (a$'$) Distribution of the magnetization for $h=0.3$ (disk radius $\rho_{\rm max}=8$). (b) Skyrmion profiles determined by Eq.~\eqref{eq:Theta}, and (c) the corresponding static magnetization determined by Eq.~\eqref{eq:m-dyn} for various magnetic fields $h<h_c$. The vertical lines are located at the radius $R_0$ of the respective skyrmion. The DMI constant $\delta=0.9$ is the same for all field values.
	}\label{fig:profiles}
\end{figure}

\subsection{General equation for static AFM skyrmion}

In Refs.~\cite{Bogdanov02a,Velkov16,Barker16,Zhang16a,Jin16a} it has been shown that the system of Eqs.~\eqref{eq:L} has a static skyrmion solution. 
Introducing the polar frame of reference for the 2D radius-vector 
$\vec{\varrho} = \rho(\cos\chi\, \hat{\vec{x}}+\sin\chi\, \hat{\vec{y}})$  within the AFM film, see Fig.~\ref{fig:setup}, one obtains for the skyrmion solution $\phi=\chi+\phi_0$ and $\theta=\Theta(\rho)$, where the function $\Theta(\rho)$ is determined by the differential problem
\begin{equation} \label{eq:Theta}
\begin{split}
&\nabla_\rho^2\Theta-\sin\Theta\cos\Theta\left(\frac{1}{\rho^2}+1-h^2\right)+\frac{|d|}{\rho}\sin^2\Theta=0,\\
&\Theta(0)=\pi,\quad\Theta(\infty)=0.
\end{split}
\end{equation}
Here $\nabla_\rho^2$ is the radial part of the Laplacian and $\phi_0=0$ ($\phi_0=\pi$) for positive (negative) DMI $d$.  In the following we restrict ourselves to the case $d>0$, as $d<0$ is completely analogous. 
A number of skyrmion profiles numerically obtained from Eq.~\eqref{eq:Theta} are shown in Fig.~\ref{fig:profiles}.
We define the skyrmion radius $R_0$ by the condition were the N\'eel vector is inplane, i.e.\ $\Theta (R_0)= \pi/2$.

For the case $h=0$, Eq.~\eqref{eq:Theta} coincides with the well known equation for a ferromagnetic skyrmion,\cite{Rohart13,Komineas15c,Leonov16,Bogdanov94,Bogdanov89} upon substitution of the N\'eel vector to the FM magnetization order parameter. However, in the presence of a magnetic field, i.e.\ $h\neq0$, there are differences in the static solution. 
While in FM energy a magnetic field enters as a linear term, in the AFM it effectively diminishes the easy-axis anisotropy.
In addition, the magnetic field creates a non-zero magnetization 
$\vec{m}= \xi h \left(\sin^2\Theta\hat{\vec{z}} -\sin\Theta\cos\Theta \hat{\vec{\rho}}\right)$ in the inhomogeneity region of the skyrmion, see Fig.~\ref{fig:profiles}(a$'$),(c). The magnetization is localized in the vicinity of the skyrmion boundary and the $m_z$ component reaches its maximum at $R_0$.


Note that in general Eq.~\eqref{eq:Theta} can only be solved numerically~\cite{Bogdanov02a,Velkov16}. Therefore in the next part we will 
review the solution in the limit of large radius skyrmions.

\subsection{Limit of large skyrmion radius} 
\begin{figure}
	\includegraphics[width=0.9\columnwidth]{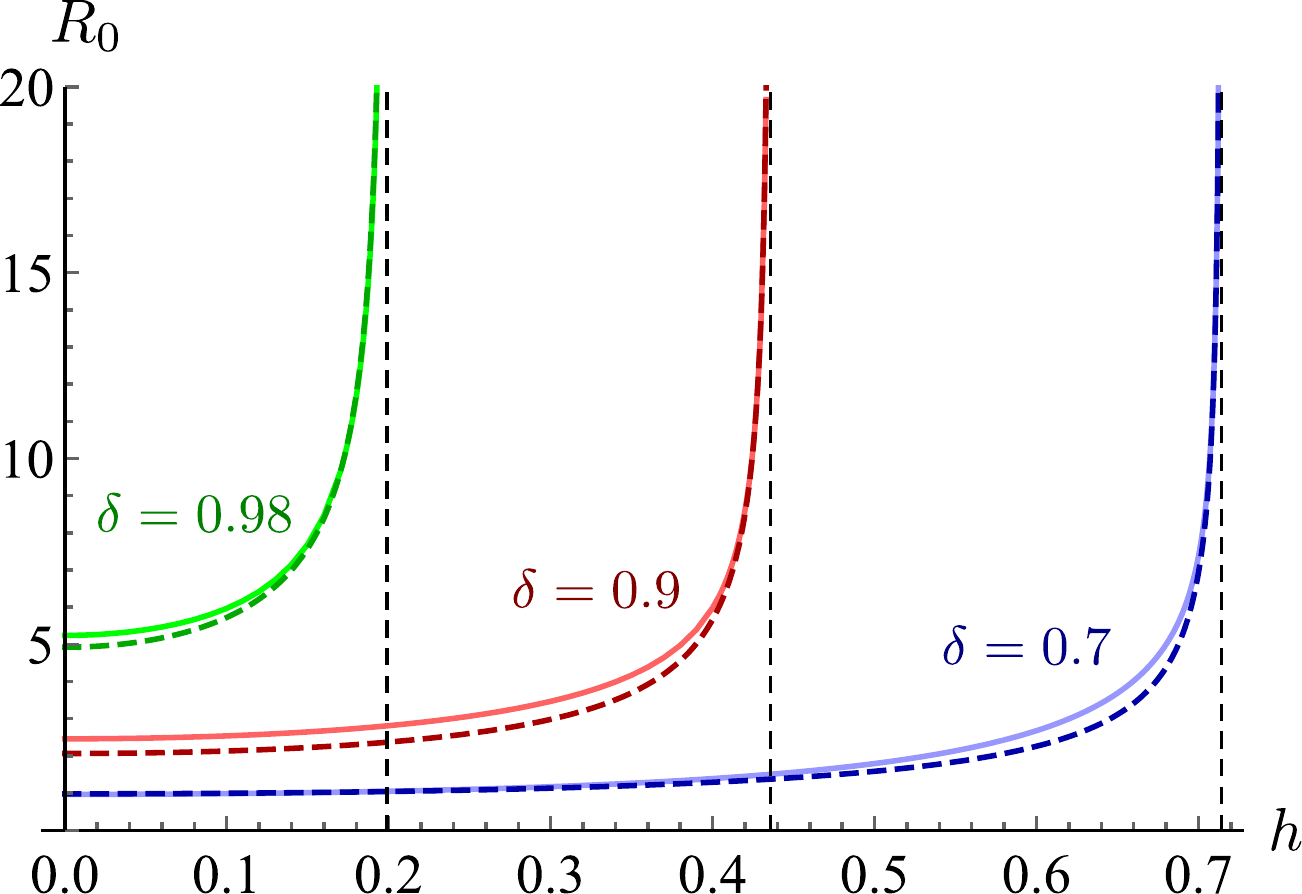}
	\caption{(Color online) \textbf{Field dependence of the skyrmion radius} for different values of DMI constants. The solid and dashed lines correspond to the numerical solution of Eq.~\eqref{eq:Theta} and the analytical approximation $R_0(h,\delta)$ determined in Eq.~\eqref{eq:R0}, respectively. The vertical asymptotes are $h_c=\sqrt{1-\delta^2}$. }\label{fig:R-vs-h}
\end{figure}

In the limit of large radius, the skyrmion can be effectively described within a few collective variables $R_0$, $\Phi_0$ and $\Delta_0$ determining the skyrmion radius, helicity, and the DW width, respectively. Transferring the knowledge of FM skyrmions \cite{Kravchuk18} and using that the magnetic field plays the role of an effective anisotropy in the AFM, the AFM skyrmion can be viewed as a circularly closed DW which can be described by
\begin{equation} \label{eq:Ansatz}
\cos\theta (\vec{\varrho}) =\tanh\frac{\rho-R_0}{\Delta_0},\quad \phi(\vec{\varrho}) =\chi+\Phi_0, 
\end{equation}
with $\Phi_0=0$, and 
 \begin{equation} \label{eq:R0}
R_0=\frac{\delta}{\sqrt{1-h^2}\sqrt{1-\delta^2-h^2}}, \quad  \Delta_0=\frac{\delta}{1-h^2}.
\end{equation}
The DW Ansatz agrees well with the exact solution of Eq.~\eqref{eq:Theta} for $R_0\gg1$, see Fig.~\ref{fig:R-vs-h} showing the field dependence of the skyrmion radius for different DMI strengths. As expected, the skyrmion radius diverges for the magnetic field reaching the critical value $h_c$.
Note that for $h=0$ the equilibrium skyrmion radius $R_0$ is given by the same expression as the radius of the FM skyrmion \cite{Rohart13, Kravchuk18}. 
The dependence of skyrmion radius on the DMI constant for $h=0$ was studied in several previous works, see e.g. Refs.~\onlinecite{Bogdanov02a,Rohart13,Kravchuk18}.

Note, that the magnetic field induces a non-zero total magnetization in a skyrmion, $\vec{\mathfrak{M}}=\!\!\int\!\!\vec{m}\,\mathrm{d}^2x$, which is aligned along the field direction. In the limit of a large-radius skyrmion and a field applied along the $z$ direction, it is $\mathfrak{M}_z=4\pi \xi h R_0\Delta_0$. This means that the static magnetic susceptibility of an AFM skyrmion is proportional to the area ($\propto 2\pi R_0\Delta_0$) of the skyrmion wall. 


\section{Eigenmodes of an antiferromagnetic skyrmion}
\label{sec:magnons}

\begin{figure}
	\includegraphics[width=0.7\columnwidth]{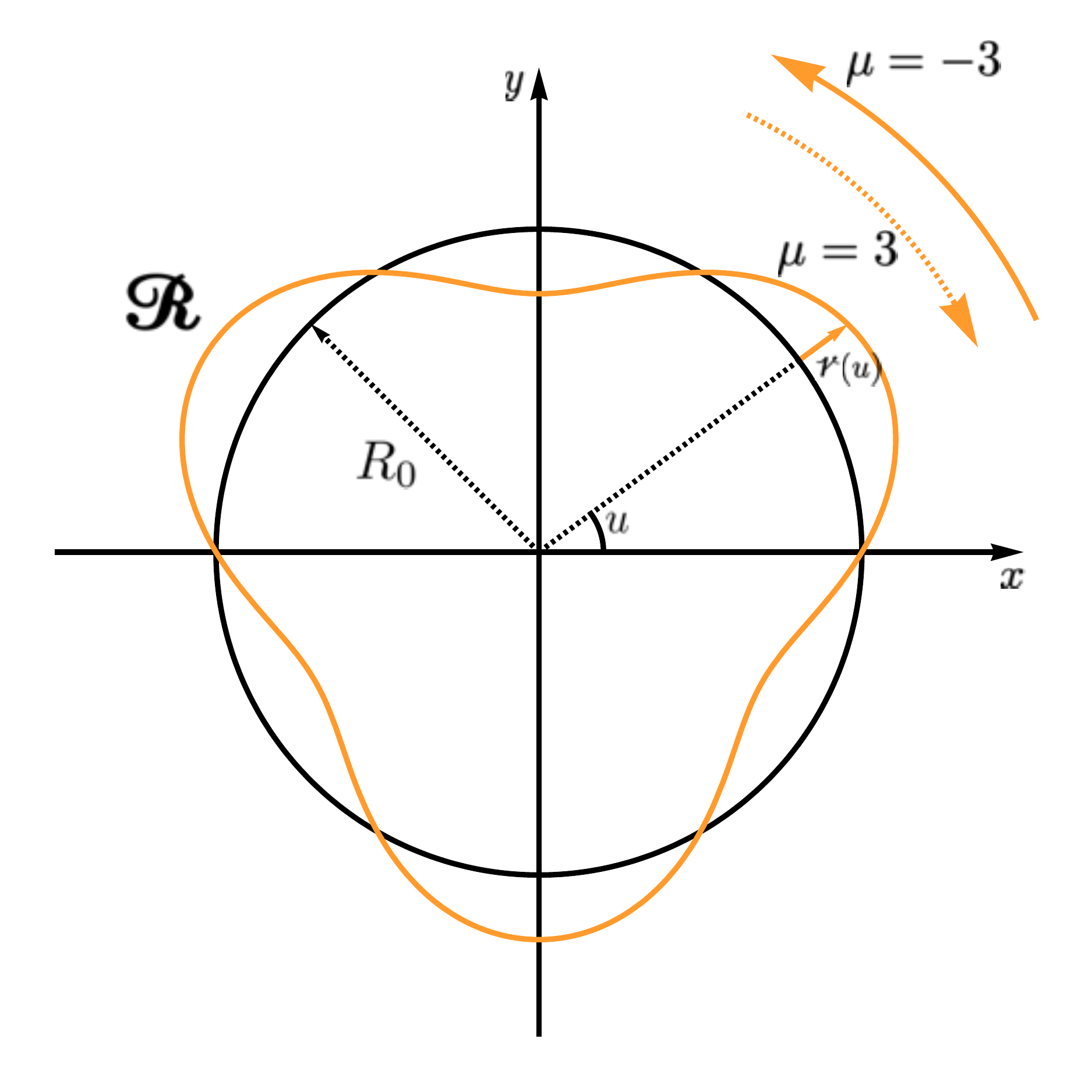}
	\caption{(Color online) {\textbf{Schematics of the perturbation}} (orange line) around the circular skyrmion configuration (black line) for $|\mu|=3$. For positive (negative) $\mu$ the rotation is clock-wise (counter clock-wise).
	}
	\label{fig:flower-like}
\end{figure}

In this section we formulate and solve numerically the exact eigenvalue  problem (EVP) for an AFM skyrmion. To this end we use a standard technique previously applied for a number of two-dimensional topological magnetic solitons, including the precessional solitons in easy-axis magnets \cite{Sheka01,Ivanov05b,Sheka06}, magnetic vortices in easy-plane magnets \cite{Ivanov98,Sheka04}, and ferromagnetic skyrmions \cite{Schuette14,Schroeter15,Kravchuk18}.
To study small excitations of AFM skyrmions, we introduce time-dependent deviations from the skyrmion solution determined by Eq.~\eqref{eq:Theta}
\begin{subequations}\label{eq:deviations}
\begin{align}
\theta&=\Theta(\rho)+\epsilon \vartheta(\rho,\chi,\tau),
\\ 
\phi&=\chi+ \epsilon \varphi(\rho,\chi,\tau)/\sin\Theta(\rho),
\end{align}
\end{subequations}
 where $\epsilon \ll 1$ and we have chosen a convenient way to accommodate for the change in the azimuthal angle. 
 Introducing above ansatz into the Lagrange function 
$\mathscr{L}$ of Eq.~\eqref{eq:L-angles} leads to the expansion $\mathscr{L}=\mathscr{L}^{(0)}+\epsilon^2\mathscr{L}^{(2)}+\dots$, where
\begin{eqnarray}
\label{eq:Lagrangeapprox}
\mathscr{L}^{(2)}&=&\dot{\vartheta}^2+\dot{\varphi}^2+V(\rho)\left(\dot{\vartheta}\varphi-\dot{\varphi}\vartheta\right)-(\nabla\vartheta)^2-(\nabla\varphi)^2\notag\\
-&U_1&(\rho)\vartheta^2-U_2(\rho)\varphi^2+W(\rho)\left(\varphi\partial_{\chi}\vartheta-\vartheta\partial_{\chi}\varphi\right).
\end{eqnarray} 
Here we introduced the potentials
\begin{equation}\label{eq:potentials}
\begin{split}
U_1(\rho)=&\cos2\Theta\left(1+\frac{1}{\rho^2}-h^2\right)-\frac{d}{\rho}\sin2\Theta,\\ 
U_2(\rho)=&\cos^2\Theta\left(1+\frac{1}{\rho^2}-h^2\right)-(\partial_{\rho}\Theta)^2\\
&-d\left(\partial_{\rho}\Theta+\frac{\sin\Theta\cos\Theta}{\rho}\right),\\ 
W(\rho)=&\frac{2}{\rho^2}\cos\Theta-\frac{d}{\rho}\sin\Theta,\\ 
V(\rho)=&2h\cos\Theta.
\end{split}
\end{equation}
In the case of $h=0$ the above potentials derived for the AFM skyrmion can be mapped fully to the ferromagnetic case \cite{Kravchuk18}. For $h\neq 0$ there appears an additional potential $V(\rho)$.
Next, we derive the equations of motion generated by the Lagrangian~\eqref{eq:Lagrangeapprox}:
\begin{equation} \label{eq:theta-phi-lin}
	\left\{
\begin{aligned}
&\ddot{\vartheta}+V\dot\varphi=\nabla^2\vartheta-U_1\vartheta-W\partial_\chi\varphi,\\
&\ddot{\varphi}-V\dot{\vartheta}=\nabla^2\varphi-U_2\varphi+W\partial_\chi\vartheta.
\end{aligned}\right.
\end{equation}
The cyclic variable $\chi$ in combination with the periodic boundary condition [$\vartheta(\rho,0,\tau)=\vartheta(\rho,2 \pi,\tau)$ and $\varphi(\rho,0,\tau)=\varphi(\rho,2 \pi,\tau)$], allows to solve Eqs.~\eqref{eq:theta-phi-lin} with the partial solutions
\begin{equation} \label{eq:product_anzatz}
\begin{split}
\vartheta & =f(\rho)\cos(\omega\tau+\mu\chi+\eta),\\
 \varphi & = g(\rho)\sin(\omega\tau+\mu\chi+\eta).
\end{split}
 \end{equation} 
 Here $\omega$ is the frequency of the corresponding eigenmode, $\mu\in\mathbb{Z}$ is the azimuthal quantum number, and $\eta$ is an arbitrary phase. Superposition of all possible partial solutions \eqref{eq:product_anzatz} composes the Fourier series of the general solution of Eq.~\eqref{eq:theta-phi-lin}. The partial solution \eqref{eq:product_anzatz} displays flower-like excitations as shown in Fig.~\ref{fig:flower-like}.
The sign of $\mu$ encodes the spatial rotation direction, $\mu>0$ being clock-wise (CW) and $\mu<0$ being counter clock-wise (CCW) with respect to the $z$-axis. 
  Substituting this ansatz into Eq.~\eqref{eq:theta-phi-lin} leads to the following eigenvalue problem (EVP)
 \begin{subequations}\label{eq:eigen-h}
	\begin{align}\label{eq:eigen-h-up}
\hat{\mathbb{H}}\vec{\Psi}=\omega\hat{\sigma}\vec{\Psi},\quad\vec{\Psi}=(f, \bar f, \bar g,g)^{\textsc{t}},
	\end{align}
	where 
	\begin{align}
\hat{\mathbb{H}}=\begin{pmatrix}
	\hat{\mathrm{H}}_1 & 0 & V & \mu W\\
	0 & 1 & 0 & 0\\
	0 & 0 & 1 & 0\\
	\mu W & V & 0 & \hat{\mathrm{H}}_2
\end{pmatrix},\quad \hat{\sigma}=\begin{pmatrix}
	0 & 1 & 0 & 0\\
	1 & 0 & 0 & 0\\
	0 & 0 & 0 & 1\\
	0 & 0 & 1 & 0
\end{pmatrix}
	\end{align}	
\end{subequations}
and $\hat{\mathrm{H}}_{1,2}=-\nabla^2_\rho+\left(U_{1,2}+\mu^2/\rho^2\right)$. Solving the EVP \eqref{eq:eigen-h-up} one finds that $\bar f= \omega f$ and $\bar g= \omega g$.

The symmetry of the Lagrangian \eqref{eq:Lagrangeapprox} with respect to translations along time and the $\chi$ coordinate results in the conservation of total energy $E=E(\mu,\omega)$ and the angular momentum $\vec{K}=\vec{K}(\mu,\omega)$, respectively. For details see App.~\ref{app:IM}.

According to Eq.~\eqref{eq:m-dyn}, the excitation of a magnon mode generates a total magnetic moment in the form 
\begin{equation}\label{eq:M-tot}
\begin{split}
\vec{\mathfrak{M}}=\vec{\mathfrak{M}}^{(0)}+&\epsilon\left[\delta_{0\mu}\vec{\mathfrak{m}}_0^{(1)}(\tau)+\delta_{1|\mu|}\vec{\mathfrak{m}}_1^{(1)}(\tau)\right]\\
+&\epsilon^2\left[\vec{\mathfrak{M}}^{(2)}_{\mu}+\delta_{0\mu}\vec{\mathfrak{m}}_0^{(2)}(\tau)\right]+\dots,
\end{split}
\end{equation}
 where $\vec{\mathfrak{M}}^{(0)}=2\pi h\xi\hat{\vec{z}}\int_0^\infty\!\rho\sin^2\Theta\,\mathrm{d}\rho$
 is the field-induced static magnetic moment discussed in Sec.~\ref{sec:static}. The time dependent parts $\vec{\mathfrak{m}}_0^{(i)}(\tau)$ and $\vec{\mathfrak{m}}_1^{(1)}(\tau)$ are linearly polarized along $\hat{\vec{z}}$-axis and circularly polarized within the $xy$-plane, respectively. Thus, the modes with $\mu=0$ and $\mu=\pm1$ (except translational mode) can be resonantly excited by the external ac field of the corresponding polarization, see App.~\ref{app:M} for details. Note that the time-independent static contribution $\vec{\mathfrak{M}}^{(2)}_{\mu}$ does not vanish even for the case $h=0$, see Eq.~\eqref{eq:M2}.

\subsection{Localized modes of an antiferromagnetic skyrmion}

\begin{figure}
	\includegraphics[width=\columnwidth]{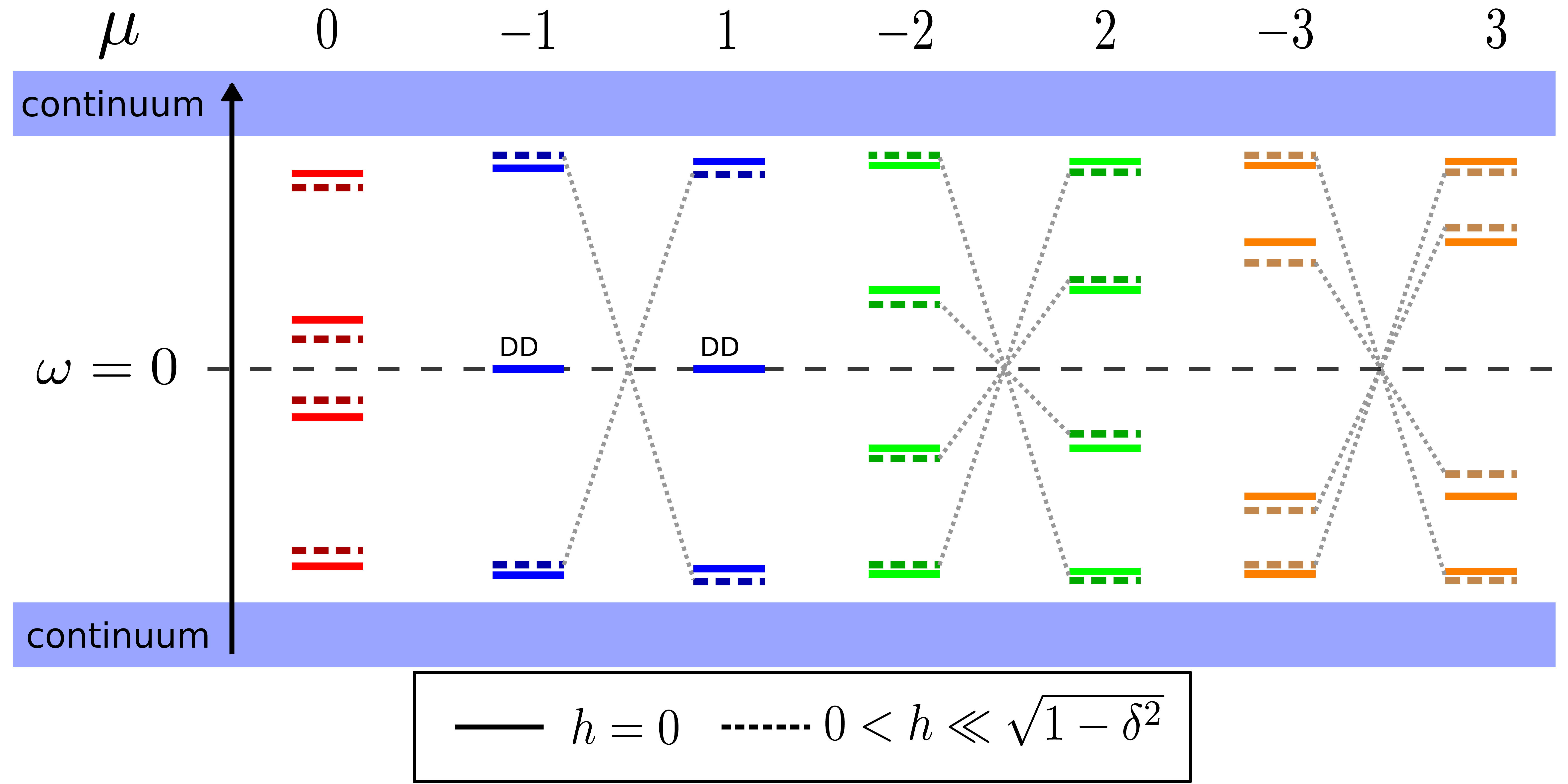}
	\caption{(Color online) \textbf{Schematics of the spectrum}. Classification of the modes with frequency $\omega_\mu$ according to their quantum number $\mu$. Gray dashed lines connect the symmetry related resonances. The modes with zero frequency and $\mu=\pm1$ are doubly degenerate (DD). 
	The field-induced narrowing of the gap (between the blue zones) is not shown. 
	}
	\label{fig:spectrum-scheme}
\end{figure}

We classify all solutions $\vec{\Psi}_{\omega_\mu}$ of the above EVP by two parameters, the frequency $\omega$ and the azimuthal quantum number $\mu$.
Due to the combined time-reversal and space-reversal symmetry of the equilibrium solution, the eigenmodes should be invariant with respect to the operations $\omega\to-\omega$, $\mu\to-\mu$, $g\to-g$ and $f\to f$ (see Eq. (\ref{eq:eigen-h})). 
Hence, the spectrum can be divided into two equivalent subsets, with negative and positive frequencies, respectively. 
The mode with zero frequency corresponds to the translational mode and will be discussed separately. 
This classification is shown schematically in Fig.~\ref{fig:spectrum-scheme}.
Following the previous studies, we will work with the subset of positive $\omega$.

\begin{figure*}
	\includegraphics[width=\textwidth]{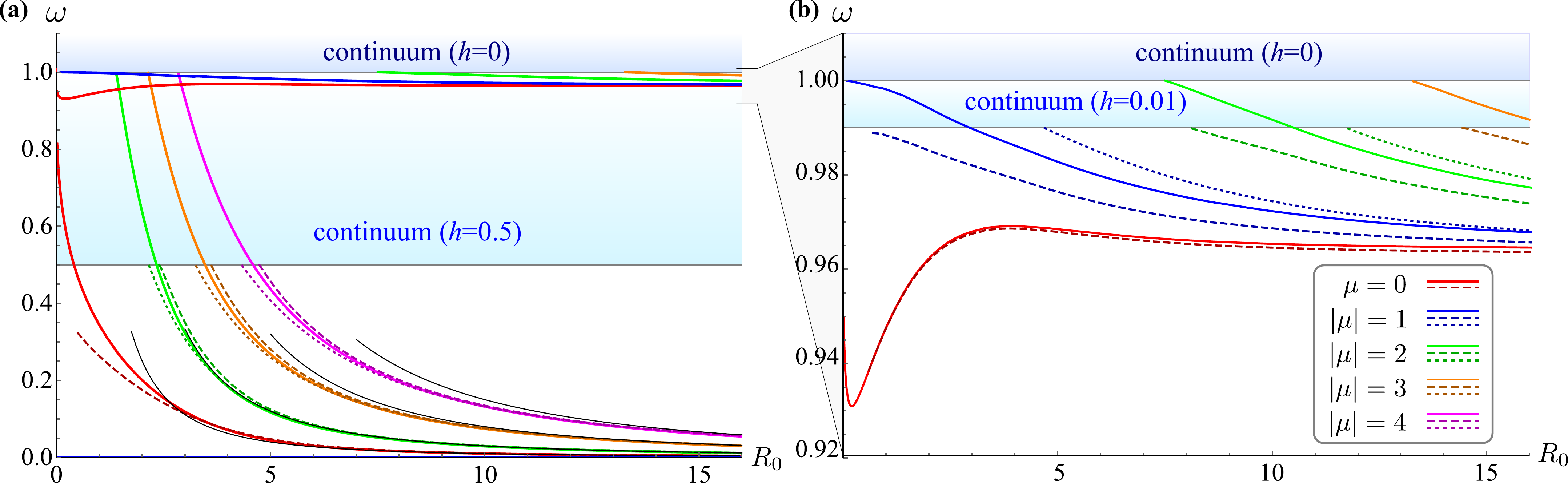}
	\caption{(Color online) \textbf{Spectrum of localized eigen-modes of an AFM skyrmion.}
	(a) spectrum for $h=0$ and $h=0.5$, (b) high frequency modes for $h=0$ and $h=0.01$ obtained by numerically solving the EVP \eqref{eq:eigen-h} with $\delta$ running the interval $[\delta_{\rm ini}(h),0.998]$ where $\delta_{\rm ini}(0)=0.08$, $\delta_{\rm ini}(0.01)=0.59$, $\delta_{\rm ini}(0.5)=0.45$.\footnote{The lower limit $\delta_{\rm ini}$ is caused by technical reasons.} The modes with $|\mu|>4$ are not shown.
	 Solid lines show the eigen-frequencies for $h=0$. Dashed and dotted lines show the eigen-frequencies of the modes $\mu\ge0$ and $\mu<0$, respectively, in the presence of a magnetic field. Thin black lines show the asymptotic approximation \eqref{eq:disp} of the LFB modes.
	}
	\label{fig:spectrum}
\end{figure*}

\begin{figure*}
\includegraphics[width=\textwidth]{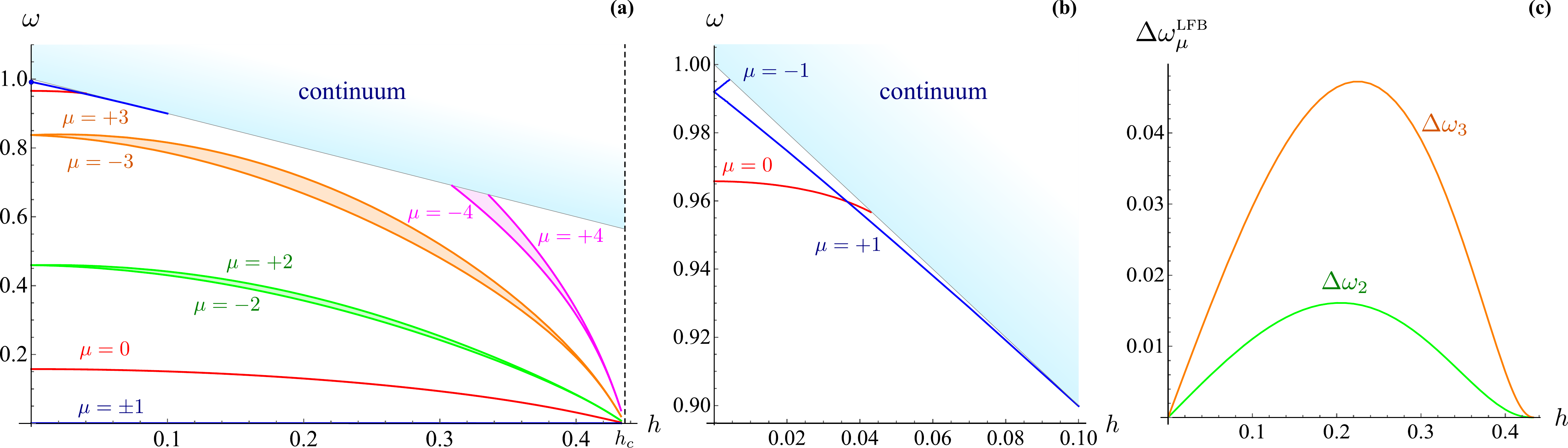}
\caption{(Color online) \textbf{Influence of the magnetic field on spectrum of the localized modes.} Field dependence of the eigen-frequencies is shown in panel (a), and HFB is detailed in panel (b). Panel (c) shows the value of splitting for the low frequency modes $\mu=\pm2$ and $\mu=\pm3$, where $\Delta\omega^{\mathrm{LFB}}_\mu=\omega_{|\mu|}-\omega_{-|\mu|}$. Here $\delta=0.9$ is fixed and $0\le h<h_c=\sqrt{1-\delta^2}$. The spectrum is obtained by means of the numerical solution of EVP \eqref{eq:eigen-h}. Note that $\mu\to-\mu$ when $h\to-h$. 
 }
\label{fig:w-h}
\end{figure*}

Our main results for the eigenmodes are shown in Figs.~\ref{fig:spectrum} and \ref{fig:w-h}
representing the spectrum of localized modes as a function of skyrmion radius and magnetic field, respectively.
We find two different types of modes which we classify as low (LFB) and a high (HFB) frequency branches. 
This can be best seen in Fig.~\ref{fig:spectrum} where all frequencies of the modes comprised in the LFB converge to zero in the large radius skyrmion limit, $\omega_\mu\propto R_0^{-2}$, see Section~\ref{sec:circ-domain-wall}, while those of the HFB do not.
Furthermore, Figs.~\ref{fig:w-h} shows that all the modes of the LFB soften at the same condition $h^2+\delta^2=1$, which corresponds to Eq.~\eqref{eq:exits}.
 The HFB modes are compactly situated at the edge of the continuum part of the spectrum.
 As a function of the magnetic field they merge into the continuum and as a function of the skyrmion radius they converge to the finite value.
 This explains also why the number of localized HFB modes increases for larger skyrmions, see Fig.~\ref{fig:spectrum}b. 
 The simultaneous softening of all LFB modes is analogous to the case of the ferromagnetic skyrmion \cite{Kravchuk18} and can be interpreted as an essential instability associated with the possibility of a continuous (soft) transition to a new state. 
However, there is no analog to the HFB in the ferromagnetic case.
 
As the magnetic field breaks time reversal symmetry, it induces a difference between CW and CCW rotational modes. This results in a splitting of the modes for $|\mu|\ge2$ in the LFB and for modes $|\mu|\ge1$ in the HFB, as explained in Figs.~\ref{fig:spectrum-scheme} and \ref{fig:w-h}.

\begin{figure*}
	\includegraphics[width=\textwidth]{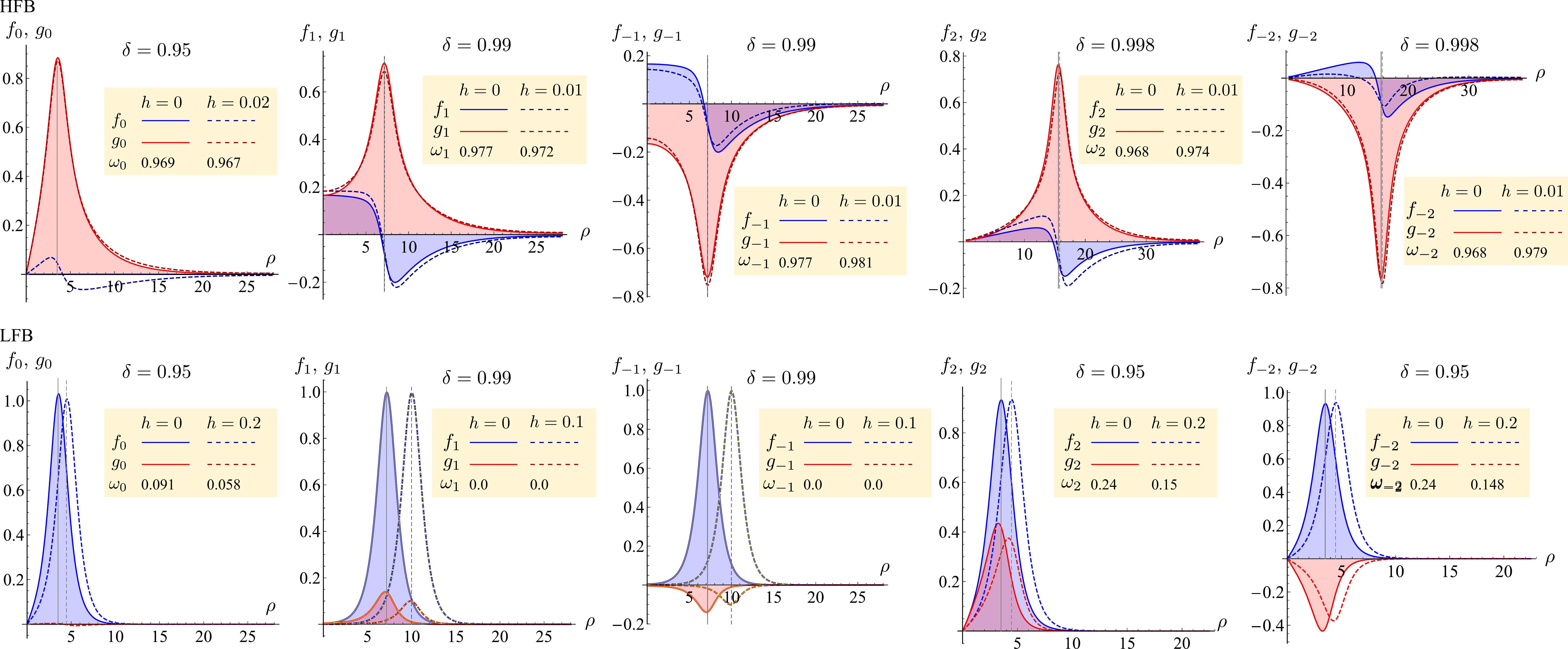}
	\caption{(Color online) \textbf{Examples of eigen-functions of localized eigen-modes} for $|\mu|\le2$. The upper and lower raws correspond to HFB  an LFB, respectively. The presented eigen-functions and eigen-frequencies are obtained by means of numerical solution of EVP \eqref{eq:eigen-h}. The vertical lines indicate the skyrmion radius $\rho=R_0$. The thin yellow lines shown for modes $\mu=\pm1$ of LFB correspond to the eigen-functions of the  translational modes: $f_{\pm1}=-\Theta'$ and $g_{\pm1}=\pm\sin\Theta/\rho$, when the normalization is applied.  
Note that in the absence of the field, $f_0\equiv0$ for the HFB (the in-plane component is only excited) and $g_0\equiv0$ for the LFB (the out-of-plane component is only excited). However, both in-plane and out-of-plane components are excited for modes with $|\mu|>0$ or in presence of a field.}
\label{fig:fg}
\end{figure*}

The eigenfunctions of the corresponding eigenmodes for different values of DMI and magnetic field separated in the HFB and LFB are shown in  Fig.~\ref{fig:fg}. 
The presented functions are normalized by the rule 
$1/2\int\limits_{0}^{\infty}\left[f_\mu^2(\rho)+g_\mu^2(\rho)\right]\mathrm{d}\rho=1$.
As a general statement, the modes in the LFB (HFB) address mainly the spherical angle $\vartheta$  ($\varphi$) of the N\'eel vector which is associated with a position change (thickness change) of the skyrmion boundary. 

In the following we will discuss the specific properties of the localized modes ordered by their quantum number $\mu$.

\begin{figure}
	\includegraphics[width=\linewidth]{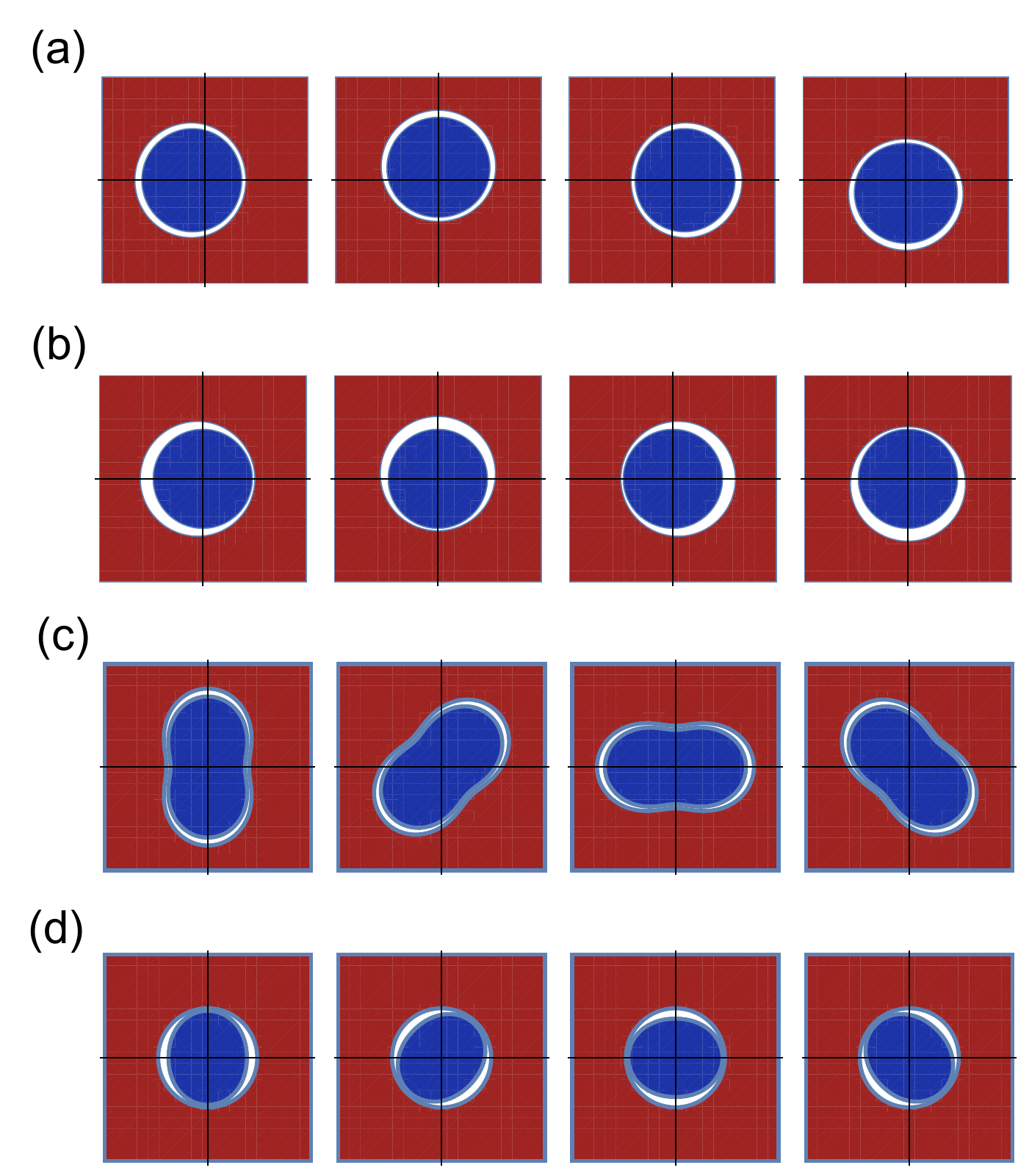}
	\caption{(Color online) \textbf{Cartoons of localized eigen-modes} for a) $|\mu|=1$ LFB showing the gyration mode with a minor effect on the domain wall thickness, b) $|\mu|=1$ HFB higher order gyration mode, where the skyrmion wall profile changes, c) $|\mu|=2$ LFB shape deformation, d) $|\mu|=2$ HFB elliptical deformation of the wall width. The color code corresponds to the out-of-plane component of the N\'eel vector. For better visibility, the amplitudes of the corrections have been enhanced.
	}\label{fig:modescartoon}
\end{figure}

\subsubsection{Radially symmetrical modes, $\mu=0$}

The localized modes with $\mu=0$ do not break the space inversion symmetry, see Fig.~\ref{fig:fg} first column.
As such, they respect the symmetry of the equilibrium solution of the skyrmion. 
Furthermore, an applied magnetic field does not couple to their dynamical degrees of freedom, see Fig.~\ref{fig:w-h}, but just changes the static solution of the skyrmion which can be translated to an effective anisotropy change. 
This explains that an increase in magnetic field leads to a reduction of their frequencies, see Figs.~\ref{fig:spectrum} and \ref{fig:w-h}.

The LFB localized mode is called breathing mode \cite{Mochizuki2012a} and corresponds to dynamics of the out-of-plane components of the N\'eel vector. It describes a dynamical contraction and expansion of the skyrmion area with the frequency $\omega_0^{\mathrm{LFB}}\approx1/R_0^2-h^2/\delta$ (see App.~\ref{sec:breathing_mode}).

The HFB mode corresponds to an in-plane oscillation of the N\'eel vector parameterized by the spherical angle $\phi$. 
 Furthermore, this mode has a total non-zero dynamic magnetization 
$\vec{\mathfrak{m}}_0^{(1)}(\tau)\propto \xi\omega_0\cos(\omega_0\tau  + \eta_0) \hat{\vec{z}}$ originated in the 
solid-like rotation of the skyrmion boundary (see App.~\ref{app:M} for the details). Therefore, this mode couples to the perpendicular ac magnetic field and can be excited resonantly, similar to breathing modes in a ferromagnetic skyrmion~\cite{Mochizuki2012a}.
For the case $h=0$ the LFB breathing mode does not generate the total dynamical moment, see Eq.~\eqref{eq:M-tot} and App.~\ref{app:M}. However, in the presence of a static magnetic field which naturally intermixes rotations and radius oscillations, the antiferromagnetic LFB breathing mode can be excited in a similar way.

Note, that decoupling of out-of-plane (LFB mode) and in-plane (HFB mode) oscillations of the N\'eel vector in absence of the magnetic field is a particular feature of an AFM skyrmion. In ferromagnetic skyrmions dynamics of the in-plane and out-of-plane components of magnetization are always coupled while  the magnetization precesses around the effective field. 

\subsubsection{Gyrotropic modes, $\mu=\pm 1$}

There are four localized modes with $|\mu|=1$, where each two of them belong to the LFB and HFB respectively. 
The modes $\mu=\pm1$ in the LFB are translational zero modes ($\omega_{\pm1}=0$) \cite{Raja1987} with the corresponding eigenfunctions $f_{\pm1}=-\partial_{\rho}\Theta$ and $g_{\pm1}=\pm\sin\Theta/\rho$.
These modes are not affected by a magnetic field as its excitation does not break the translational invariance. 
 In contrast to the ferromagnetic skyrmion \cite{Kravchuk18,Satywali18}, an AFM skyrmion has two translational modes which correspond to two independent degrees of freedom related to the motion of the skyrmion center.  
In analogy to light, where two linearly polarized beams can be used to create circular polarized light, these two modes can be combined to CW and CCW gyrotropic modes with zero frequency, see Fig.~\ref{fig:modescartoon}.  
The translational modes do not create a dynamic magnetization (see App.~\ref{app:M}).

The modes in the HFB with $\omega_{\pm1}\neq 0$ are different gyrotropic modes where the skyrmion wall width oscillates, see Fig.~\ref{fig:modescartoon}. This is reflected by the fact that the in-plane $g$-component is more intensive than the out-of-plane $f$-component, which has a node in vicinity of the skyrmion radius, see Fig.~\ref{fig:fg}. This also means that the degeneracy is lifted by a magnetic field, Figs.~\ref{fig:spectrum} and~\ref{fig:w-h}.  In contrast to the translational modes, the HFB counterparts generate magnetic moment $\vec{\mathfrak{m}}_1^{(1)}(\tau)$ even in absence of the external magnetic field. The moment $\vec{\mathfrak{m}}_1^{(1)}$ uniformly rotates within the $xy$-plane, see \eqref{eq:m_mu1}. So, this mode can be resonantly exited by the circularly polarized ac magnetic field.

\subsubsection{Higher modes, $|\mu|\ge2$}

\hyphenation{con-ti-nuum}

For $|\mu|\ge2$ the modes appear below the continuum only when the skyrmion radius exceeds a certain $\mu$-dependent threshold.
The LFB modes resemble the flower-like modes similar to the FM case. \cite{Sheka01,Makhfudz12, Lin2014, Chen2018a, Kravchuk18} 
The HFB modes are related with skyrmion wall width changes. 
All the modes split due to a magnetic field. 
For the LFB, the value of the field-induced frequency splitting is non-monotonic in field, has a maximum for $0<h<h_c$, and vanishes 
when the magnetic field approaches the critical field, $h\to h_c$, see Fig.~\ref{fig:w-h}(c). 
This can be understood intuitively as follows. An increase of the field leads to a reduction in the anisotropy, thereby increasing the skyrmion radius. When approaching to the critical field, the skyrmion radius diverges, see Fig.~\ref{fig:R-vs-h}. In this case, the central part of the skyrmion resembles an AFM with vanishing magnetization and effective anisotropy. This effectively restores the time reversal invariance and the degeneracy of the two modes.
Note, that the modes with $|\mu|\ge2$ cannot be excited by spatially homogeneous fields.

In the following section we will explain our results within an intuitive analytical model.


\section{Simple model of the antiferromagnetic domain wall string}
\label{sec:circ-domain-wall}

The aim of this section is to obtain an analytical estimation for the eigenfrequencies. We focus on the LFB modes and consider the limit $R_0\gg1$. In this case the AFM skyrmion can be considered as a circularly closed AFM domain wall. Since the field influence on the LFB modes vanishes for the large-radius skyrmions, see Fig.~\ref{fig:spectrum}(a), we restrict the discussion to the field-free case.
For the modes in the LFB we observe that the main effect of the excitation is to modify the shape of the skyrmion boundary.
As the excitations are localized sharply on the skyrmion boundary (see Fig.~\ref{fig:fg}) and leave the skyrmion wall width almost unchanged, (see Figs.~\ref{fig:modescartoon}a,c) in the following we will treat the skyrmion eigenmode dynamics in terms of an effective string model accounting only for the position of the skyrmion boundary and its curvature.
Recently, the analogous formalism was developed for ferromagnetic domain walls \cite{Zhang18c,Rodrigues18} and it was applied for the description of large radius skyrmion eigenmodes. Here the skyrmion boundary was treated as a curve or string representing the effective coordinate of the system.

\subsection{Curvilinear Lagrange formalism for antiferromagnetic N\'eel domain walls}

Here we consider a 2D antiferromagnetic  DW, whose center position, can be described by a curve
$\vec{\mathcal{R}}(u,\tau)=\hat{\vec{x}} \mathcal{X}(u,\tau) + \hat{\vec{y}} \mathcal{Y}(u,\tau)$, 
where $u$ is the time independent parameter of the curve and $\tau$ describes a potential time dependence, see Fig.~\ref{fig:flower-like}.
 Along the curve $\vec{\mathcal{R}}$ one can introduce a local orthonormal basis 
\begin{equation} \label{eq:basis}
	\vec{e}_\textsc{t}(u,\tau)=\frac{\vec{\mathcal{R}}'}{|\vec{\mathcal{R}}'|},\qquad \vec{e}_\textsc{n}(u,\tau)=\frac{\hat{\vec{z}}\times\vec{\mathcal{R}}'}{|\vec{\mathcal{R}}'|},
\end{equation}
where prime denotes the derivative with respect to $u$. The vectors $\vec{e}_\textsc{t}$, $\vec{e}_\textsc{n}$ and $\hat{\vec{z}}$ are the tangential, normal and binormal vectors for the Frenet-Serret basis of the curve, respectively~\cite{Kuehnel15}.
In vicinity of the curve $\vec{\mathcal{R}}$ the position of any point $(x,y)$ can be determined by a pair of local curvilinear coordinates $(\mathrm{u},\mathrm{v})$, which are defined as a solution of the following vector equation
	\begin{equation} \label{eq:curvilinear-frame}
	\vec{r}=\vec{\mathcal{R}}(\mathrm{u},\tau) + \mathrm{v}\vec{e}_{\textsc{n}}(\mathrm{u},\tau),
	\end{equation}
	where $\vec{r}=x\hat{\vec{x}}+y\hat{\vec{y}}$.
	Note, that due to the explicit time dependence of the vectors $\vec{\mathcal{R}}$ and $\vec{e}_{\textsc{n}}$ the curvilinear coordinates $(\mathrm{u},\mathrm{v})$ are time-dependent \footnote{Note, there is difference between $u$ and $\mathrm{u}$: $u$ is time-independent parameter of the curve, while $\mathrm{u}$ is a time-dependent coordinate  of a point in the $xy$-plane.}.
In formal terms, we require $|\mathrm{v}\varkappa|\ll1$, where $\varkappa=\hat{\vec{z}}\cdot\left[\vec{\mathcal{R}}'\times\vec{\mathcal{R}}''\right]/|\vec{\mathcal{R}}'|^3$ 
is the curvature of $\vec{\mathcal{R}}$.
This representation allows us to parametrize the N\'eel vector at any point in the vicinity of the curve $\vec{\mathcal{R}}$.

Motivated by the LFB eigenfunctions (shown in Fig.~\ref{fig:fg}), where the dominant contribution of the skyrmion excitation is related with the angle $\theta$ (described by $f_{\mu}$), we restrict ourselves to the following N\'eel wall ansatz \footnote{For a circularly closed DW the normal vector $\vec{e}_\textsc{n}$ is oriented inwards as it is used in curvilinear approaches.
Therefore, the signs ``$-$'' appears in Eq.~\eqref{eq:str-Ansatz}.} 
\begin{subequations} \label{eq:str-Ansatz}
\begin{align} \label{eq:str-Ansatz-1}
\vec{n} &= \cos\varTheta \hat{\vec{z}} - \sin\varTheta \vec{e}_\textsc{n},\\
\label{eq:str-Ansatz-2}
\cos\varTheta &= -\tanh\frac{\mathrm{v}(\tau)}{\Delta}.
\end{align}
\end{subequations}

The action for our system is given by 
\begin{subequations} \label{eq:action}
\begin{align} \label{eq:action-1}
\mathcal{S} &= \int \!\! \mathrm{d}\tau \! \int \!\! \mathrm{d}S \mathscr{L} = 2
\int \!\! \mathrm{d}\tau \! \int \!\! \mathrm{d}\mathrm{u}\,  \mathscr{L}^{\mathrm{eff}},
\end{align}
\end{subequations}
where the Lagrange function $\mathscr{L}$ is given by Eq.~\eqref{eq:L} and
 $\mathrm{d}S=|\vec{\mathcal{R}}'|(1-\varkappa \mathrm{v})\mathrm{d} \mathrm{u}\mathrm{d}\mathrm{v}$ is the infinitesimal area element in curvilinear coordinates. 
After substituting our ansatz (Eqs.~\eqref{eq:str-Ansatz}) into the Lagrangian and integrating out the normal coordinate $\mathrm{v}$ we obtain the effective Lagrangian $\mathscr{L}^{\mathrm{eff}}$ for the domain wall string, for details see App.~\ref{app:DW-string}:
\begin{subequations} \label{eq:L-afm-smpl}
\begin{align} \label{eq:L-afm-smpl-L}
\!\!\!
\mathscr{L}^{\mathrm{eff}} &= \mathscr{T}^\mathrm{eff} - \mathscr{E}^\mathrm{eff},\\
\label{eq:L-afm-smpl-T} 
\!\!\!
\mathscr{T}^\mathrm{eff} &= \frac{\left[\dot{\vec{\mathcal{R}}}\times\vec{\mathcal{R}}'\right]^2}{\Delta|\vec{\mathcal{R}}'|} + \frac{\Delta\left[\vec{\mathcal{R}}'\times\dot{\vec{\mathcal{R}}}'\right]^2}{|\vec{\mathcal{R}}'|^3},\\
\label{eq:L-afm-smpl-E} 
\!\!\!
\mathscr{E}^{\mathrm{eff}} &= \frac{|\vec{\mathcal{R}}'|}{\Delta}\Biggl[ \! 1 - 2\delta \Delta + \Delta^2\left(1 + \varkappa^2\right) + \mathcal{O}\left(\varkappa^4\Delta^4 \right) \! \Biggr]\!.
\end{align}
\end{subequations}
The effective Lagrangian is a sum of a pure kinetic term $\mathscr{T}^\mathrm{eff}$ and the potential energy of the domain wall $\mathscr{E}^{\mathrm{eff}}$ originating from the different contributions in Eq.~\eqref{eq:L-angles}.
This Lagrange formalism provides the basis for the description of the slow dynamics of the AFM N\'eel domain walls and skyrmions with small curvatures.
 In the next subsection we will apply it to the particular case of small excitations of circular AFM skyrmions. For further illustration of the formalism we discuss the simplest example, i.e.\ the one of an AFM domain wall in App.~\ref{sec:dw}.

\subsection{Circular AFM skyrmion}
\label{sec:circular-wall}

In general, a large radius skyrmion can be represented as a circular domain wall $\vec{\mathcal{R}} = R \left(\hat{\vec{x}} \cos\varPhi + \hat{\vec{y}}\sin\varPhi\right)$, where the angle $\Phi$ is the parameter of the curve ($0\leq \Phi \leq 2\pi$) and $R$ is the radius of the skyrmion. 
For the model of Eq.~\eqref{eq:L-afm-smpl} the static equilibrium solution is given by $R_0=\delta/\sqrt{1-\delta^2}$, and $\Delta_0=\delta$, which corresponds to the circular DW coinciding with our previous results, Eq.~\eqref{eq:R0}. 
To describe excitations, we introduce a small deviation from the static solution, $R= R_0 +  \mathcal{r}(u,\tau)$ and $\varPhi=u + \psi(u,\tau)$. With this one can derive the
harmonic part of the effective Lagrangian \eqref{eq:L-afm-smpl} as follows:
\begin{equation} \label{eq:L-eff-CDW}
\mathscr{L}^\mathrm{eff} \approx R_0^2(R_0^2+1)\dot{\mathcal{r}}^2 + R_0^2(R_0\dot{\psi}-\dot{\mathcal{r}}')^2-{\mathcal{r}}''^2 + 2{\mathcal{r}}'^2 - {\mathcal{r}}^2.
\end{equation}
As the effective Lagrangian does not depend explicitly on $\psi$ (cyclic variable), the time independent quantity $\mathcal{j}=R_0\dot{\psi}-\dot{\mathcal{r}}'$ is conserved. $\mathcal{j}$ plays the role of an orbital momentum corresponding to a rigid rotation of the system and can be set to be zero in the inertial frame.
Thus, we obtain the Euler-Lagrange equation for the radius 
\begin{equation} \label{eq:r}
R_0^2(R_0^2+1)\ddot{\mathcal{r}}+\mathcal{r}^{(\textsc{iv})}+2\mathcal{r}''+\mathcal{r}=0.
\end{equation}

The latter has solutions in the form of azimuthal waves, $\mathcal{r}(u,\tau) =\mathcal{r}_0 \cos(\omega_\mu\tau+\mu u+\eta)$
with dispersion
\begin{equation} \label{eq:disp}
\omega_{\mu}=\frac{|1-\mu^2|}{R_0\sqrt{R_0^2+1}}\approx\frac{|1-\mu^2|}{R_0^2}.
\end{equation}
For the deviation of the angle we then find $\psi(u,\tau) =\psi_0 \sin(\omega_\mu\tau+\mu u+\eta)$, where $\psi_0 = -\mathcal{r}_0 \mu/R_0$. For the breathing mode $\mu=0$ the amplitude $\psi_0=0$ of the in-plane oscillations vanishes. This is in a full agreement with the numerical results shown in Fig.~\ref{fig:fg} and  with the analytical prediction \eqref{eq:R-Phi-lin}.
Note that Eq.~\eqref{eq:disp} also correctly describes the translational modes with $\mu=\pm1$ and $\omega_{\pm1}=0$.

The asymptotic solutions of $\omega_{\mu}$ presented by \eqref{eq:disp} are shown in Fig.~\ref{fig:spectrum}(a) by thin solid black lines.


\section{Conclusions}
\label{sec:concl}

We demonstrated that the AFM skyrmion has a discrete spectrum of bound eigenstates. The properties of these states were studied both analytically and numerically. The spectrum consists of two branches. The modes of the low-frequency branch demonstrate a significant dependence on the skyrmion radius, see Eq.~\eqref{eq:disp}, and they are responsible for the skyrmion instability when the DMI strength exceeds some critical value. The modes of the high-frequency branch are compactly situated at the magnon continuum and they are not involved in the instability process. All the modes, except the radially symmetrical one, are doubly degenerated with respect of the sense of rotation around the skyrmion center: clockwise or counterclockwise. An out-of-plane magnetic field removes the degeneracy (for all modes except translational), resulting in a frequency splitting, which for the small fields is linear in field. For the large-radius skyrmion the low-frequency modes can be interpreted as oscillations of a circularly closed AFM domain wall (geometric degree of freedom). The high-frequency modes corresponds to oscillations of the magnetic deviations from the domain wall structure (magnetic degree of freedom). The external magnetic field mixes the geometrical and magnetic degrees of freedom.
The high-frequency modes with $\mu=0$ and $\mu=\pm1$ can be excited by the ac magnetic field which is linearly polarized along $z$-axis or circularly polarized within $xy$-plane, respectively. The low-frequency breathing mode ($\mu=0$) can be excited by the perpendicular ac field only in the presence of an external bias field along $z$-axis. The translational modes can not be excited by the magnetic fields.

In addition to our numerical study, we have developed a formalism to describe extended AFM domain walls in terms of a collective string dynamics. Applying our theory to AFM skyrmion  interpreted as circular AFM domain
walls, we analytically predict the asymptotic for the low frequency part of the AFM skyrmion spectrum.

Overall our study of the AFM skyrmion excitations, provides guidance for experiments to detect them. For example, we expect that, in particular, for large radius skyrmions the low energy excitations can be observed by BLS techniques.

\begin{acknowledgements}
V.P.K. thanks Ulrike Nitzsche for technical assistance. D.D.S. acknowledges the support from the Alexander von Humboldt Foundation (Research Group Linkage Programme) and Taras Shevchenko National University of Kyiv (project 19BF052-01). This work was supported by National Academy of Sciences of Ukraine, Project No. 0116U003192 and by  UKRATOP project funded by the German Federal Ministry of Education and Research, Grant No. 01DK18002. K.E.-S, D.R.R. and O.G. acknowledge funding of the Transregional Collaborative Research Center (SFB/TRR) 173 SPIN+X. K.E.-S. acknowledges funding from the German Research Foundation (DFG) under the Project No. EV 196/2-1. O.G. and J.S. acknowledge funding from the Humboldt Foundation,  the ERC Synergy Grant SC2 (No. 610115), the EU FET Open RIA Grant no. 766566, the DFG (project SHARP 397322108).

\end{acknowledgements}

\appendix


\section{Details of the DW string model}
\label{app:DW-string}

In this appendix we show how to derive the effective Lagragian $\mathscr{L}^{\mathrm{eff}}$ in Eq.~\eqref{eq:action} of the main text
\begin{subequations} \label{eq:action_A}
\begin{align} 
\label{eq:action-2} %
\mathscr{L}^{\mathrm{eff}} &= \frac{|\vec{\mathcal{R}}'|}{2} \int \!\! \left(1-\varkappa \mathrm{v}\right) \mathscr{L} \mathrm{d}\mathrm{v} = \mathscr{T}^{\mathrm{eff}} - \mathscr{E}^{\mathrm{eff}},
\end{align}
\end{subequations} 
where the kinetic and potential energy parts using Eq.~\eqref{eq:L} are given by
\begin{subequations} \label{eq:TandEeff}
\begin{align} 
\label{eq:Teff} %
\mathscr{T}^{\mathrm{eff}} &= \frac{|\vec{\mathcal{R}}'|}{2} \int \!\! \left(1-\varkappa \mathrm{v}\right)
\dot{\vec{n}}^2  \mathrm{d}\mathrm{v} ,\\
\label{eq:Veff}
\mathscr{E}^{\mathrm{eff}} &= \frac{|\vec{\mathcal{R}}'|}{2} \int \!\! \left(1-\varkappa \mathrm{v}\right) \mathcal{W} \mathrm{d}\mathrm{v}.
\end{align}
\end{subequations} 

Let us compute now the effective kinetic energy $\mathscr{T}^{\mathrm{eff}}$. 
For the ansatz of Eq.~\eqref{eq:str-Ansatz} we obtain $\dot{\vec{n}}^2 = \dot{\varTheta}^2 + \sin^2\varTheta \dot{\vec{e}}_{\textsc{t}}^2$ with 
$\dot{\varTheta}=\sech(\mathrm{v}/\Delta)\dot{\mathrm{v}}/\Delta$. 
Taking into account that the time derivative of a coordinate vector $\vec{r}$ is zero,  we differentiate Eq.~\eqref{eq:curvilinear-frame} by time and solve the obtained set of equations with respect to $\dot{\mathrm{v}}$ and $\dot{\mathrm{u}}$. This results in $\dot{\mathrm{v}}=\hat{\vec{z}}\cdot\left[\dot{\vec{\mathcal{R}}}\times\vec{\mathcal{R}}'\right]/|\vec{\mathcal{R}}'|$, where the $\dot{\vec{\mathcal{R}}}=\partial\vec{\mathcal{R}}/\partial\tau$ is the time derivative with respect to the explicit time dependence.  With this we obtain
\begin{equation} \label{eq:theta-dot}
\dot{\varTheta} =
\frac{\hat{\vec{z}}\cdot\left[\dot{\vec{\mathcal{R}}}\times\vec{\mathcal{R}}'\right]}{\Delta |\vec{\mathcal{R}}'|} \sech\left(\frac{\mathrm{v}}{\Delta}\right).
\end{equation}
Using the definition of the basis vectors, Eq.~\eqref{eq:basis} one obtains
\begin{equation} \label{eq:e-t-dot}
\dot{\vec{e}}_{\textsc{t}}^2=\dot{\vec{e}}_{\textsc{n}}^2 = \frac{\left[\dot{\vec{\mathcal{R}}'}\times\vec{\mathcal{R}}'\right]^2}{ |\vec{\mathcal{R}}'|^4}.
\end{equation}
By integrating \eqref{eq:Teff} we derive the effective kinetic energy of the domain wall string, Eq.~\eqref{eq:L-afm-smpl-T}.

For the derivation of the effective  potential energy of the domain wall $\mathscr{E}^{\mathrm{eff}}$ we need to calculate the defined in \eqref{eq:L-2} energy density $\mathcal{W}=\mathcal{W}_{\rm ex}+\mathcal{W}_{\rm an}+\mathcal{W}_{\textsc{dmi}}$ for the case of Ansatz \eqref{eq:str-Ansatz} formulated in the curvilinear coordinates $(\mathrm{u},\mathrm{v})$. While the anisotropy contribution is trivial $\mathcal{W}_{\rm an}=\sin^2\Theta$ the calculation of the exchange $\mathcal{W}_{\rm ex}=(\vec{\nabla}\varTheta)^2 + \vec{\varOmega}^2\sin^2\varTheta$ and DMI $\mathcal{W}_{\textsc{dmi}}=- 2\sin^2\varTheta \left(\vec{e}_{\textsc{n}} \cdot \vec{\nabla} \varTheta \right)$ terms requires the technique previously developed for the curvilinear systems  \cite{Gaididei14,Kravchuk18a}.  Here $\vec{\nabla}=\vec{e}_\textsc{t}|\vec{\mathcal{R}}'|^{-1}(1-\kappa \mathrm{v})^{-1}\partial_\mathrm{u}+\vec{e}_\textsc{n}\partial_\mathrm{v}$ is the surface gradient and $\vec{\varOmega}=-\varkappa(1-\varkappa \mathrm{v})^{-1}\vec{e}_\textsc{t}$ is the vector of spin-connection \cite{Kamien02,Bowick09}. The usage of the mentioned techniques requires the metric tensor $||g_{\alpha\beta}||=\text{diag}\left(|\vec{\mathcal{R}}'|^2(1- \varkappa \mathrm{v})^2, 1\right)$, which can be straightforwardly obtained from the parameterization \eqref{eq:curvilinear-frame} with the application of the Frenet–-Serret formulaes $\vec{e}_{\textsc{t}}'=|\vec{\mathcal{R}}'|\varkappa\,\vec{e}_\textsc{n}$ and $\vec{e}_\textsc{n}'=-|\vec{\mathcal{R}}'|\varkappa\,\vec{e}_\textsc{t}$. The area element reads $\mathrm{d}S=\sqrt{|g|}\mathrm{d} \mathrm{u}\mathrm{d}\mathrm{v}$. 

Integrating now Eq.~\eqref{eq:Veff} leads to the effective potential energy for the domain wall string 
in the form of Eq.~\eqref{eq:L-afm-smpl-E}.

\section{Domain wall string}
\label{sec:dw} %
In this section we consider the model of Eq.~\eqref{eq:L} and illustrate the application of the string model to excitations of a planar antiferromagnetic DW, which is described by the equilibrium solution $\vec{\mathcal{R}}_0=u \hat{\vec{x}}$ and $\Delta_0=1$.
Let us introduce small deviations $\vec{\mathcal{R}} = \vec{\mathcal{R}}_0 + \mathcal{X}(u,\tau) \hat{\vec{x}} + \mathcal{Y}(u,t)\hat{\vec{y}}$, where $|\mathcal{X}|, |\mathcal{Y}|\ll1$. In harmonic approximation with respect to the deviations $\mathcal{X}$ and $\mathcal{Y}$ the Lagrangian in Eq.~\eqref{eq:L-afm-smpl} is given by
\begin{equation}\label{eq:L-DW}
\mathscr{L}^\mathrm{eff}\approx \dot{\mathcal{Y}}^2+\dot{\mathcal{Y}}'^2-\mathcal{Y}''^2-(1-\delta)\mathcal{Y}'^2.
\end{equation}
The corresponding equation of motion 
\begin{equation}\label{eq:DW-dyn}
\ddot{\mathcal{Y}}-\ddot{\mathcal{Y}}''-(1-\delta)\mathcal{Y}''+\mathcal{Y}^{(\textsc{iv})}=0
\end{equation}
has a plane--wave solution $\mathcal{Y}(u,\tau)\propto \cos\left(q u - \omega \tau+\eta\right)$ with the spin--wave spectrum
\begin{equation}\label{eq:DW-disp}
\omega=|q|\sqrt{1-\delta} ,
\end{equation}
where $q$ is the wave-vector along the DW. This linear dispersion relation coincides with the results found in Ref.~\onlinecite{Ivanov1992}.
In contrast to oscillations of a ferromagnetic DW \cite{Makhfudz12,Zhang18} the excitations of the antiferromagnetic DWs in the low frequency branch do not show a nonreciprocity effect, even in the presence of DMI (opposite directions $\pm q$ are equivalent). 

\section{Modes with $\mu=0$ in large skyrmion limit}\label{sec:breathing_mode} %
In this section we derive approximate expressions for both LFB and HFB with $\mu=0$ in presence of the magnetic field, whose asymptotics is not fully covered with the string model. As in the field absence the modes with $\mu=0$ keep the symmetry of equilibrium skyrmion solution, they can be approximately described  with the same ansatz \eqref{eq:Ansatz} as equilibrium skyrmion: 
\begin{equation} \label{eq:Ansatz_A} 
\cos\theta (\vec{\varrho},\tau) =\tanh\frac{\rho-R(\tau )}{\Delta},\quad \phi(\vec{\varrho},\tau) =\chi+\Phi(\tau ),
\end{equation}
where $\Delta$  is the domain wall width.

In the absence of a magnetic field a LHB (HFB) mode with $\mu=0$  corresponds to pure out-of-plane (in-plane) oscillations of the N\'eel vector seen as the
radial (tangential) oscillations of the DW. 
 
The dynamics of the collective variables is determined by the effective Lagrange function 
\begin{equation} \label{eq:L-eff}
\mathcal{L}^{\mathrm{eff}}=\frac{R\dot{R}^2}{\Delta}+\Delta R(\dot{\Phi}-h)^2-\frac{R}{\Delta}-\frac{\Delta}{R}-\Delta R+2\delta R\cos\Phi,
\end{equation}
where we neglect possible explicit time dependence of $\Delta$.

Using symmetry arguments we represent the collective variables as $\Phi=\Phi_0+\psi(\tau)$, $R=R_0+ \mathcal{r}(\tau)$, where $\psi(\tau)$ and $\mathcal{r}(\tau)$ are small excitations of equilibrium soliton solution with the parameters $\Phi_0$ and $R_0$ given by Eq.~\eqref{eq:R0}). 
Linearized equations of motion for excitations are then obtained from the Lagrange function \eqref{eq:L-eff} as
\begin{subequations}\label{eq:R-Phi-lin}
	\begin{align}
		&\ddot{\mathcal{r}}+\omega_L^2 \mathcal{r}  +h\frac{\Delta_0^2}{R_0}\dot{\psi}=0,\\ 
		&\ddot{\psi}+\omega_H^2\psi-\frac{h}{R_0}\dot{\mathcal{r}}=0,
	\end{align}
\end{subequations}
where $\omega_L=\Delta_0/R_0^2$ and $\omega_H=\sqrt{\delta/\Delta_0}$. 

In absence of the magnetic field ($h=0$) the eigenfrequencies of LFB and HFB modes are
\begin{equation}\label{eq:w-h0}
\omega_{L0}=\frac{1-\delta^2}{\delta}=\frac{1}{R_0\sqrt{R_0^2+1}}\approx\frac{1}{R_0^2},
\end{equation}
and $\omega_{H0}=1$, respectively. 

The external magnetic field induces coupling of the radial and tangential components. The eigenfrequency of LFB mode calculated from Eqs.~\eqref{eq:R-Phi-lin}  in approximation of weak magnetic field ($h\ll1$) is 
\begin{equation}\label{eq:w-h}
\omega_0\approx\omega_{L0}-\left[\frac{\delta}{2}\frac{\omega_{L0}^2}{\omega_{H0}^2-\omega_{L0}^2}+\frac{1}{\delta}\right]h^2.
\end{equation}
Predicted field-induced frequency decrease $\propto h^2$  correlates well with the results of numerical calculations (see Fig.~\ref{fig:w-h}a).
Note, that the model of Eq.~\eqref{eq:Ansatz} does not distinguish the HFB mode from the magnon continuum and field-dependence of this mode can be studied only numerically.

\section{Integrals of motion}\label{app:IM}
The Lagrange function in Eq.~\eqref{eq:Lagrangeapprox} does not explicitly depend on spatial azimuthal angle $\chi$ and time. This leads to two integrals of motion. The first is associated to the rotation invariance around the $z$-axis and corresponds to the angular momentum  $\vec{K}=\int\left[\vec{r}\times\vec{\mathcal{P}}_\chi\right]\mathrm{d}^2x$. Here $\vec{\mathcal{P}}_\chi=\vec{e}_\chi\rho^{-1}\left[2(\dot{\vartheta}\partial_\chi\vartheta+\dot{\varphi}\partial_\chi\varphi)+V(\rho)(\varphi\partial_{\chi}\vartheta-\vartheta\partial_{\chi}\varphi)\right]$ is  density of the linear momentum corresponding to the translations along $\chi$. In terms of the partial solutions \eqref{eq:product_anzatz}, one obtains
\begin{equation} \label{eq:K}
\vec{K}=2\pi\mu\hat{\vec{z}}\int\limits_0^\infty\rho\left[\omega\left(f^2+g^2\right)-Vfg\right]\mathrm{d}\rho.
\end{equation}
It is linear in the azimuthal quantum number $\mu$. The second integral of motion is the total energy  $E=\int\mathcal{E}\mathrm{d}^2x$, where $\mathcal{E}=\dot{\vartheta}^2+\dot{\varphi}^2+(\nabla\vartheta)^2+(\nabla\varphi)^2+U_1\vartheta^2+U_2\varphi^2-W(\varphi\partial_{\chi}\vartheta-\vartheta\partial_{\chi}\varphi)$. Substitution with the partial solutions \eqref{eq:product_anzatz} provides
\begin{equation}\label{eq:E}
\begin{split}
E=\pi\int_{0}^{\infty}&\rho\biggl[\omega^2(f^2+g^2)+f'^2+g'^2+2\mu Wfg+\\
&+\left(\frac{\mu^2}{\rho^2}+U_1\right)f^2+\left(\frac{\mu^2}{\rho^2}+U_2\right)g^2\biggr]\mathrm{d}\rho.
\end{split}
\end{equation} \\
As it follows from \eqref{eq:K} and \eqref{eq:E} both integrals of motion depend on the two parameters: $\vec{K}=\vec{K}(\mu,\omega)$ and $E=E(\mu,\omega)$.

\section{Dynamical magnetic moment}\label{app:M}
Applying \eqref{eq:deviations} and \eqref{eq:product_anzatz} to 
\eqref{eq:m-dyn} and integrating over the film area one straightforwardly obtains \eqref{eq:M-tot}, where
\begin{equation}\label{eq:m_mu0}
\vec{\mathfrak{m}}_0^{(1)}(\tau)=\xi\left(\omega_0\mathcal{A}^{\omega}+h\mathcal{A}^{h}\right)\cos(\omega_0\tau  + \eta_0)\hat{\vec{z}},
\end{equation}
with $\mathcal{A}^{\omega}=-2\pi\int_{0}^{\infty}\!\!\rho g_0\sin\Theta\,\mathrm{d}\rho$ and $\mathcal{A}^{h}=2\pi\int_{0}^{\infty}\!\!\rho f_0\sin2\Theta\,\mathrm{d}\rho$.  Note that for the LFB one has $\mathcal{A}^{\omega}=0$ if $h=0$. This is because $g_0\equiv0$ for this case, see Fig.~\ref{fig:fg}. Thus, in the absence of a static field, $\vec{\mathfrak{m}}_0^{(1)}\equiv0$ for the LFB and the corresponding breathing mode can not be excited by the external field.   
\begin{equation} \label{eq:m_mu1}
\begin{aligned}
\vec{\mathfrak{m}}_1^{(1)}&(\tau)=\xi\left(\omega_{\pm1}\mathcal{B}^{\omega}_{\pm1}+h\mathcal{B}^{h}_{\pm1}\right) \\
&\times\left[\cos(\omega_{\pm1}\tau  + \eta_{\pm1})\hat{\vec{x}}\mp\sin(\omega_{\pm1}\tau  + \eta_{\pm1})\hat{\vec{y}}\right],
\end{aligned}\end{equation}
where $\mathcal{B}^{\omega}_{\pm1}=\pi\int_{0}^{\infty}\!\!\rho\left[g_{\pm1}\cos\Theta\mp f_{\pm1}\right]\mathrm{d}\rho$ and $\mathcal{B}^{h}_{\pm1}=\pi\int_{0}^{\infty}\!\!\rho\left[\pm g_{\pm1}\cos\Theta-f_{\pm1}\cos2\Theta \right]\mathrm{d}\rho$. Thus, the gyrotropic modes can be excited by the in-plane ac field of the circular polarization. However, such a field can not induce the skyrmion motion, because for the translational modes $\omega_{\pm1}=0$ and $\mathcal{B}^{h}_{\pm1}=0$. The modes with $|\mu|\ge2$ do not contribute to the linear (in $\epsilon$) part of the total magnetic moment.

For any $\mu$ there appears the static magnon induced contribution 
\begin{equation}\label{eq:M2}
\vec{\mathfrak{M}}^{(2)}_{\mu}=\pi\xi\hat{\vec{z}}\int\limits_{0}^{\infty}\rho\left[hf_{\mu}^2\cos(2\Theta)-2\omega_{\mu} f_{\mu}g_{\mu}\cos\Theta\right]\mathrm{d}\rho,
\end{equation}
which survives also for the case $h=0$. If $\mu=0$ then the quadratic (in $\epsilon$) part of the total moment gains the time dependent part $\vec{\mathfrak{m}}_0^{(1)}(\tau)=\vec{\mathfrak{M}}^{(2)}_{0}\cos(2\omega_0\tau+2\eta_0)$ with the doubled frequency.

%
%
%
%

\end{document}